\documentclass[prl,aps,showpacs,twocolumn]{revtex4}
\usepackage{amssymb}

\usepackage{graphicx}

\begin{document}

\title{Controlled DNA compaction within chromatin:
the tail-bridging effect}
\author{Frank M\"{u}hlbacher, Christian Holm}
\author{Helmut Schiessel}

\altaffiliation{Present address: Instituut-Lorentz, Universiteit
Leiden, Postbus 9506, 2300 RA Leiden, The Netherlands}

\affiliation{Max-Planck-Institut f\"{u}r Polymerforschung, Theory
Group, PO Box 3148, D 55021 Mainz, Germany}

\date{\today}

\begin{abstract}

We study the mechanism underlying the attraction between
nucleosomes, the fundamental packaging units of DNA inside the
chromatin complex. We introduce a simple model of the nucleosome,
the eight-tail colloid, consisting of a charged sphere with eight
oppositely charged, flexible, grafted chains that represent the
terminal histone tails. We demonstrate that our complexes are
attracted via the formation of chain bridges and that this
attraction can be tuned by changing the fraction of charged
monomers on the tails. This suggests a physical mechanism of
chromatin compaction where the degree of DNA condensation can be
controlled via biochemical means, namely the acetylation and
deacetylation of lysines in the histone tails.

\end{abstract}


\maketitle

In eukaryotes (plants and animals) meters of DNA have to be
compacted inside micron-sized nuclei. At the same time a
considerable fraction of the genetic code has to be accessible.
Nature has solved this formidable task by compacting DNA in a
hierarchical fashion \cite{Schiessel03}. The first step consists
of wrapping the DNA two turns around cylinders made from eight
histone proteins. This leads to a string of cylindrical DNA spools
about 10 nm in diameter and 6 nm in height, each repeating unit
being called a nucleosome \cite{Luger97}. The chromatin fiber with
diameter of about 30 nm is typically posited as the next
compaction level which again forms higher order structures such as
loops. The density of such structures varies along the fiber and
in the course of the cell-cycle and is presumably directly related
to the genetic activity with the dense regions corresponding to
"silenced" parts.

It is far from being obvious how nature copes with the challenge
of combining high compaction and (selective) accessibility at the
same time. Recently -- via the combination of experiments and
theory -- an understanding has begun to emerge of how the
nucleosome is meticulously designed to face this challenge. In
principle, when DNA is wrapped onto the protein cylinder, it is in
a "closed" state not accessible for DNA binding proteins. But
thermal fluctuations open a window of opportunity for such
proteins via the unwrapping of either one of the two turns
\cite{Polach95,Kulic04} or via a corkscrew sliding of the octamer
\textit{along} the DNA chain \cite{Gottesfeld02,Kulic03}. Also
remodelling complexes can actively induce nucleosome sliding along
DNA \cite{Becker02}.

Less clear, however, is the situation at the next levels of
compaction. The chromatin fiber has a roughly 40 times shorter
contour length than that of the DNA chain it is made from. But at
the same time the fiber is much stiffer than the naked chain, so
the coil size that would be formed by a chromatin fiber in dilute
solution would still be much larger than the diameter of the cell
nucleus \cite{footnote1}. This clearly calls for the necessity of
nucleosome-nucleosome attraction as a further means of compaction.
This mechanism should be tunable such that fractions of the fiber
are dense and transcriptionally passive, while others are more
open and active.

This leads to several important questions: Can nucleosomes attract
each other, and what, if so, is the underlying mechanism? Can this
interaction be tuned for individual nucleosomes? And can this all
be understood in simple physical terms? Recent experiments indeed
point towards a simple mechanism that leads to attraction between
nucleosomes: the histone tail bridging
\cite{Mangenot02a,Mangenot02b, Bertin04}. The histone tails are
flexible extensions of the eight core proteins that carry several
positively charged residues \cite{Luger97,Luger98}. These tails
extend considerably outside the globular part of the nucleosome.
Mangenot et al. \cite{Mangenot02a} studied dilute solutions of
nucleosome core particles (NCPs; the particles that are left when
the non-adsorbed "linker" DNA is digested away). Via small angle
X-ray scattering it was demonstrated that NCPs change their size
with increasing salt concentration: At around 50 mM monovalent
salt the radius of gyration increases slightly (from 43 \AA\ to 45
\AA), but at the same time the maximal extension of the particle
increases significantly (from 140 \AA\ to 160 \AA). This
observation was attributed to the desorption of the cationic
histone tails from the NCP, which carries an overall negative
charge (cf.~Ref.~\cite{Schiessel03}). Osmometric measurements
\cite{Mangenot02b} detected around the salt concentration where
the tails desorb an attractive contribution to the interaction
between the NCPs, manifest in a considerable drop of the second
virial coefficient. The coincidence of the ionic strengths for the
two effects led Mangenot et al. to suggest that it is the tails
that are mainly involved in the attractive interaction
\cite{footnote2}. This picture was recently supported by another
study \cite{Bertin04} where it was shown that the attraction
disappeared after the tails on the NCPs had been removed.

On the theoretical side the role of histone tails is not clear.
Attraction between simplified model nucleosomes has been observed
\cite{Beard01,Boroudjerdi03}, yet both models ignore the tails. In
the former study \cite{Beard01} the NCP crystal structure has been
mimicked by a cylinder with 277 charge patches accounting for
charged groups on the surface of the NCP (without tails). In the
latter study \cite{Boroudjerdi03} the nucleosome was modelled by a
negatively charged sphere and a semiflexible cationic chain
wrapped around. The interaction between two such complexes (in the
ground state approximation) shows an attraction at intermediate
salt concentrations leading to a non-monotonic behavior of the
second virial coefficient (cf.~Fig.~4 in \cite{Boroudjerdi03}).
The only theoretical study focusing on tail bridging was recently
presented by Podgornik \cite{Podgornik_03}. The NCP was modelled
by a point-like particle with an oppositely charged flexible chain.
This system showed NCP-NCP attraction but there was no
non-monotonic behavior of the second virial coefficient. This
leads to the question: Is tail bridging responsible for the
attraction between NCPs observed at intermediate salt
concentrations? Or is this attraction rather based on correlations
between charged patches \cite{Rouzina96}? The latter possibility
is supported by a recent computer simulation of Allahyarov et al.
\cite{Allahyarov03} who studied the interaction between spherical
model proteins that carry charge patches; the second virial
coefficient featured a non-monotonic behavior as a function of
ionic strength. Also the non-monotonic interaction found by
Boroudjerdi and Netz \cite{Boroudjerdi03} belongs to that class of
attraction induced by charge correlations \cite{Rouzina96}.

The purpose of the present study is fourfold: (\textit{i}) to
introduce a minimal model for an NCP including its tails,
(\textit{ii}) to test whether such a model shows attraction with a
non-monotonically varying  second virial coefficient,
(\textit{iii}) to put tail bridging on a stronger footing and
demonstrate that this effect is qualitatively different from
attraction through charge patches, and (\textit{iv}) to
demonstrate how tail bridging can be used to facilitate control of
the compaction state of chromatin.

\begin{figure}[tbp]
\includegraphics*[width=7cm]{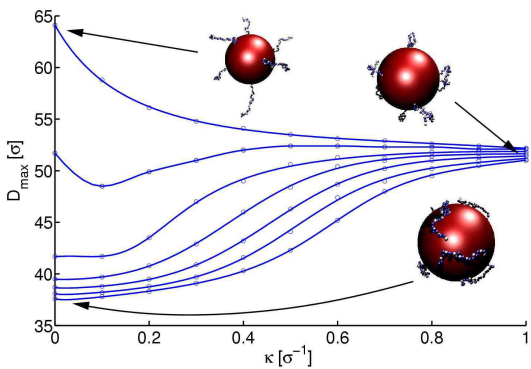}
\caption{Average maximal extension of the eight-tail colloid as a
function of the salt concentration together with three example
configurations. The different curves correspond to different
values of the central charge: $\left|Z\right|=0, 50, 100, 150,
200, 250, 300$ (from top to bottom).} \label{fig1}
\end{figure}

We start with presenting our NCP model, the eight-tail colloid
depicted in Fig.~\ref{fig1}. It consists of a sphere with eight
attached polymer chains. The sphere is a coarse-grained
representation of the NCP without the tails, i.e., the globular
protein core with the DNA wrapped around. The sphere carries a
central charge $Z$ that represents the net charge of the
DNA-octamer complex; since the DNA overcharges the cationic
protein core, one has $Z<0$ \cite{Schiessel03}. Furthermore, the
sphere radius is chosen to be $a=15 \sigma$ with $\sigma=3.5$ \AA\
being our unit length. The eight histone tails are modelled by
flexible chains grafted onto the sphere (at the vertices of a
cube). Each chain consists of 28 monomers of size $\sigma$ where
each third monomer carries a positive unit charge, the rest being
neutral. All these parameters have been chosen to match closely
the values of the NCP~\cite{footnote3}. The simulations were
performed in a NVT ensemble, using a Langevin thermostat
\cite{frenkel02b} with a time step of 0.01 $\tau$, and a friction
coefficient $\Gamma = \tau^{-1}$ (Lennard-Jones time unit). The
hard cores were modelled with a WCA potential, the chain
connectivity with a FENE potential, and the central sphere was
allowed to freely rotate (cf. Ref.~\cite{frank_preprint} for
details of the implementation). In addition, all charged monomers
and the central sphere experience an electrostatic interaction via
the standard Debye-H\"uckel (DH) theory with an inverse screening
length $\kappa=\sqrt{4\pi l_B c_s}$, where $c_s$ denotes the
monovalent salt concentration and $l_B= 2 \sigma$ sets the Bjerrum
length in water at room temperature ($l_B=e^2/\epsilon k_B T$:
$e$: electron charge, $\epsilon$: dielectric constant of solvent,
$k_B T$: thermal energy) \cite{mcquarrie}. Since we use a DH
potential, we need to use an effective value $Z_{\text{eff}}$ for
the central charge to account for charge renormalization
\cite{Alexander84}.

Figure \ref{fig1} presents results of a Molecular Dynamics (MD) simulation of
a single eight-tail colloid. Depicted is the thermally averaged maximal
extension of the colloid as a function of $\kappa$ for different values of the
central sphere charge $Z$. For $Z=0$ and small values of $\kappa$, i.e., at
low ionic strength, the eight tails are extended, radially pointing away from
the center of the complex, cf. the example at $\kappa \sigma=0$. For large
values of $\left|Z\right|$, say, for $\left|Z\right|>100$, and small $\kappa$
the tails are condensed onto the sphere, cf.~the configuration at
$\left|Z\right|=300$ and $\kappa \sigma=0$.  Increasing the screening leads in
both cases finally to structures where the chains form random polymer coils as
the ones in the example at $\kappa \sigma=1$. With increasing values of
$\left|Z\right|$ the swelling of initially condensed tails sets in at larger
$\kappa$-values. A comparison of our curves for $\left|Z\right|>100$ with the
experimental ones \cite{Mangenot02a} shows a qualitatively similar chain
unfolding scenario.  Furthermore, by choosing $Z=-150$ we are able to match
closely the experimental and the simulation values of $c_s$ at which tail
unfolding takes place. In the following we will therefore always use this
value as our $Z_{\text{eff}}$.

\begin{figure}[tbp]
\includegraphics*[width=7cm]{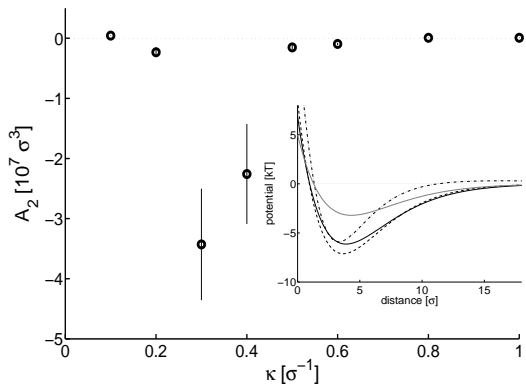}
\caption{Second virial coefficient of the eight-tail colloid as a
function of salt concentration. The inset shows the interaction
potential between two eight-tail colloids as a function of the
surface-surface separation for 4 different values of $\kappa$: $
\kappa \sigma=0.2$ (dashed-dotted line), $\kappa \sigma=0.3$
(dashed line), $ \kappa  \sigma=0.4$ (solid line) and $\kappa
\sigma=0.6$ (grey line).} \label{fig2}
\end{figure}

We determined next the interaction between two such complexes by
measuring the thermally averaged force at different distances and
by interpolating the force-distance curve via a suitable
least-square fit. Integration then yields the pair potentials
depicted in the inset of Fig.~\ref{fig2} for four different values
of $\kappa$. We find an attractive potential with a minimum of a
few $k_B T$ in all four cases. The depth of the potential shows a
non-monotonic dependence on $\kappa$ with a maximal value around
$\kappa \sigma=0.3$. This in turn is reflected in a non-monotonic
dependence of the second virial coefficient $A_2$
(cf.~Fig.~\ref{fig2}) with a minimum around the $\kappa$-value
where tail unfolding occurs, cf.~the curve for
$Z=Z_{\text{eff}}=-150$ in Fig.~\ref{fig1}. Again, all these
observations are qualitatively similar to the experimental ones
\cite{Mangenot02b}.

\begin{figure}[tbp]
\includegraphics*[width=7cm]{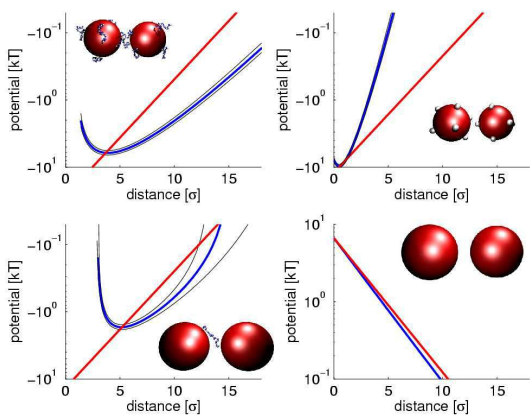}
\caption{Comparison of the interaction potential (with error
corridor) for 4 different colloids at $\kappa \sigma=0.4$:
eight-tail colloids (top left), colloids with charge patches (top
right), one-tail bridging (bottom left) and homogeneously charged
balls (bottom right). For each model we depict the potential in a
semilogarithmic plot (only the attractive part for the three first
cases). The curves are compared to a line with slope $\pm
\kappa$.} \label{fig3}
\end{figure}

Having a simulation model at hand allows us now to determine
whether this attraction can be attributed to the tail-bridging
effect. In Fig.~\ref{fig3} we depict a comparison of the full
eight-tail model with simplified variants. In all cases we choose
$\kappa \sigma=0.4$, a value close to the one where $A_2$ has its
minimal value in Fig.~\ref{fig2}; $\kappa \sigma=0.4$ corresponds
to 100 mM monovalent salt, i.e., to physiological conditions. In
one case (top right) we collapse each chain on a small patch
modelled as a grafted monomer that carries the whole chain charge.
Inspecting the attractive part of the pair potential we see that
this patch model has a very rapidly decaying interaction with a
slope larger than the reference line with slope $\kappa$. In sharp
contrast, the eight-tail complex has a decay constant that is
smaller than $\kappa$ (cf.~top left of Fig.~\ref{fig3}), an effect
that can only be attributed to tail bridging. This effect can also
be seen for our third variant (bottom left) where 15 of the 16
tails have been removed and $Z$ has been adjusted so that the net
charges of the complexes are unchanged. The remaining one-tail
complex is not allowed to rotate so that the grafting point of the
chain always faces the other ball. Also in that case the range of
attraction is longer than expected from pure screened
electrostatics. Finally, on the bottom right we present the
trivial case of two charged balls (with the same net charge as the
full model) where only a \textit{repulsive} interaction remains.

\begin{figure}[tbp]
\includegraphics*[width=7cm]{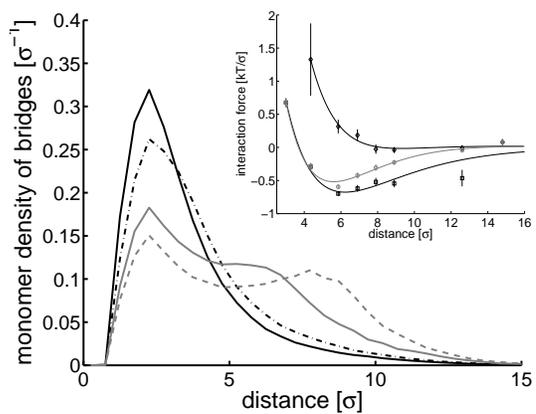}
\caption{Density distribution of monomers belonging to
bridge-forming tails as a function of the distance from the
surface of the colloid to which the tail is grafted. The different
distributions correspond to different surface-surface separations
between colloids: $d=0\sigma$ (solid), $d=4\sigma$
(dashed-dotted), $d=7 \sigma$ (grey) and $d=9 \sigma$ (dashed).
The inset separates the total average of the interaction force
(circles) into the part stemming from configurations with bridges
(squares) and non-bridging configurations (diamonds).}
\label{fig4}
\end{figure}

Having established the qualitative difference between tail-induced
attraction and attraction via charge patches we take in
Fig.~\ref{fig4} a closer look at the tail-bridging effect between
two eight-tail colloids, again for $\kappa \sigma=0.4$. Depicted
is the monomer distribution of bridge-forming chains. We define
such a chain as a chain that has at least one of its monomers
closer than a distance $3.6 \sigma$ to the surface of the alien
core. For very small distances between the colloids there are
almost always bridges. Their monomer distribution shows a strong
peak around a distance $3 \sigma$. However, also at much larger
distances like $d=7 \sigma$ and $d=9 \sigma$ there is still a
considerable fraction of configurations that show bridges. Their
monomer distribution shows a bimodal distribution with the two
peaks clearly reflecting the condensation of monomers on the home
core and the alien core. The inset shows the interaction force
between two colloids (circles) and the contributions of
tail-bridging configurations (squares) and configurations without
bridges (diamonds) to this force. It can be clearly seen that the
tail-bridging configurations account to an overall attractive
force, whereas in the other case the interaction is on average
purely repulsive.

\begin{figure}[tbp]
\includegraphics*[width=6cm]{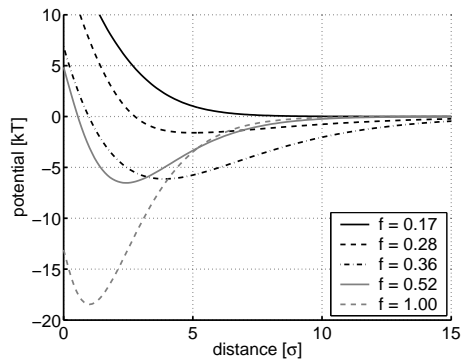}
\caption{Interaction potential between two eight-tail complexes as
a function of the surface-surface separation for $\kappa
\sigma=0.4$ and various charge fractions $f$.} \label{fig5}
\end{figure}

Finally, we speculate how the tail bridging can be used by the
cellular machinery to control DNA compaction and genetic activity.
We have determined the pair potential between eight-tail complexes
for different charge fractions of the tails. As can be seen in
Fig.~\ref{fig5}, its equilibrium distance goes to larger values
and finally disappears when one goes from a charge fraction
$f=0.36$ (the value used above) to $f=0.17$. It is in fact known
that the cellular machinery is capable of controlling the charge
state of the histone tails via the acetylation (the "discharging")
and deacetylation (the "charging") of its lysine groups
\cite{Horn02}. Active, acetylated regions in chromatin are more
open, inactive, deacetylated regions tend to condense locally and
on larger scales as well \cite{Tse98}. For instance, chromatin
fibers tend to form hairpin configurations once a sufficiently
strong internucleosomal attraction has been reached
\cite{Mergell04,Grigoryev99}. This suggests a biochemical means by
which the degree of chromatin compaction and genetic activity can
be controlled via a physical mechanism, the tail-bridging effect.

{\textit{Acknowledgement}: The authors thank M. Deserno, B.
D\"unweg, K. Kremer, F. Livolant, S. Mangenot and R. Podgornik for
helpful discussions.


\begin{thebibliography}{99}
\bibitem{footnote1} The size of a stiff polymer chain in a good
solvent scales like $R\approx l_P^{1/5}D^{1/5}L^{3/5}$ ($l_P$:
persistence length, $D$: diameter, $L$ contour length)
\cite{Odijk78}. A human chromosomal DNA chain has $L\approx 4$ cm.
This together with $l_P=50$ nm and $D \approx 4$ nm (assuming
physiological ionic conditions) leads to $R\approx
100\mu\mbox{m}$. On the other hand the chromatin fiber has $L
\approx 1$ mm, $l_P \approx 200$ nm \cite{Mergell04} and $D
\approx 30$ nm leading to $R\approx 20\mu\mbox{m}$. Note that
there are 46 chains that have to fit into the nucleus with a
diameter of 3 to 10 $\mu\mbox{m}$.
\bibitem{footnote2} Note the opposite effect of colloidal stabilization
that occurs when polyelectrolyte chains are grafted densely onto
colloids, cf.~\cite{Pincus91}.
\bibitem{footnote3} We choose here the average length of the N-terminal tails
whose lengths range from 15 residues (histone H2A) to 44 (H3). The
tail bridging effect reported here is very robust: variations in
e.g. the anchor positions have only a minor impact on the
interaction between our complexes \cite{frank_preprint}. In the
context of the cell some tails (the two H3 tails) might be
involved in the interaction with the linker DNA thereby
controlling the DNA entry-exit angle as detailed in chapter 3.5 of
Ref.~\cite{Schiessel03}.
\end{thebibliography}
\end{document}